# EVOLUTION OF FREE VOLUME ELEMENTS IN AMORPHOUS POLYMERS UNDERGOING UNIAXIAL DEFORMATION: A MOLECULAR DYNAMICS SIMULATION STUDY

Brendan Wernisch[*], Mohammed Al Otmi[*], Egan Beauvais, Janani Sampath

* These authors contributed equally to this work.

Department of Chemical Engineering, University of Florida, Gainesville, Florida 32611

jsampath@ufl.edu

**ABSTRACT**

Amorphous polymers are considered promising materials for separations due to their excellent transport properties and low fabrication costs. The separation performance of a membrane material is characterized by its permeability (overall throughput of components), and selectivity (efficiency of separation). Both permeability and selectivity are controlled by the diffusion of different penetrants through the matrix, which is strongly influenced by the distribution and morphology of the free volume elements (FVEs). FVEs are void spaces in the polymer matrix that result from the inefficient packing of bulky and rigid groups on the polymer backbone. Thus, FVEs dictate the efficiency of membrane polymers, and it is imperative to understand how processing conditions such as high pressure influence their structure. In this paper, we apply uniaxial tensile deformation on three polymers, namely polystyrene (PS), polymethylpentene (PMP), and HAB-6FDA thermally rearranged polymer (TRP), at varying temperatures and strain rates. We calculate the stress strain curve, tensile modulus, and free volume element evolution at these conditions. We find that PMP and PS with low and moderate glass transition temperature, respectively, exhibit the most change in mechanical properties as a function of strain rate and temperature. The properties of TRP, however, do not vary as much. We also find that FVEs become larger with deformation, and the extent of this change is in line with the overall change of mechanical properties of the material.

# INTRODUCTION

Energy intensive processes, such as cryogenic distillation, are currently used to separate hydrocarbon mixtures in petrochemical industries, accounting for a significant fraction of the industry's energy consumption [1]–[3]. Specifically, cryogenic distillation is implemented to obtain high purity olefins for plastic manufacturing from olefin/paraffin mixtures [4]. In this regard, polymer membranes offer a more efficient and sustainable alternative to reduce the thermal energy requirement in hydrocarbon separations by up to 90% [5], [6]. Amorphous polymer membranes, in particular, offer many advantages including scalability, processability, cost-effective fabrication, and tunable transport and mechanical properties [7]. Beyond olefin/paraffin separations, amorphous polymers have been applied in several other industrial separation applications, including water desalination and purification [8], [9], natural gas production [10], [11], carbon capture [12], [13], and fuel cells [14], [15].

The performance of a gas separation membrane is measured by its permeability and selectivity [16]; an ideal membrane is both highly permeable and highly selective. However, these two qualities are inversely correlated, and this is best demonstrated by the Robeson plot of membrane selectivity versus permeability. The so-called "upper bound" of this plot describes the inherent tradeoff between permeability and selectivity, where increased permeability often leads to decreased selectivity[17], [18]. Transport in these membranes can be described using the solution-diffusion model, which defines permeability as a product of diffusivity and solubility [19]. In rigid polymers, diffusivity is the dominant contributor to a membranes' transport properties compared to solubility. Particularly, diffusivity is influenced by the pore distribution in the polymer matrix, which are not discrete void spaces, but well-connected unoccupied volume. This unoccupied volume is also known as free volume elements (FVE), which form as a result of the inefficient packing of bulky groups along the polymer backbone and side chains [20]. As a gas mixture diffuses through a membrane, its transport is facilitated by FVEs. It is for this reason that FVE distribution is an integral parameter to tune while designing membrane materials, as it controls transport properties of the membrane module.

While gas transport influences membrane efficiency, mechanical properties dictate membrane lifetime. Understanding mechanical response of membrane materials under different conditions is crucial for optimizing the durability and investment cost of a membrane module. In addition to having superior transport properties, membrane materials must have high strength to withstand high-pressure operations necessary to drive molecular transport. At the same time, membrane materials need to be flexible enough to be safely implemented in myriad design modules without undergoing fracture. As the membrane experiences structural deformation, the sizes and distribution of its FVEs change, impacting the membrane's transport properties. Additionally, it is important to understand the influence of temperature on mechanical properties as well, as many separation processes occur over a temperature range. Temperature plays an important role in influencing the materials' deformation responses, impacting its mechanical strength and the nature of FVE evolution during deformation. The overarching goal is to design appropriately flexible membrane modules with robust, durable FVE distributions, which can maintain performance at different operating temperatures.

Studies have investigated membrane polymers' response to structural deformation [21]–[23], and some related this analysis with an interdependent examination of the effect on FVE morphology, FVE evolution, and the effects of deformation on FVE-facilitated transport properties [24]–[26]. One study by K. Emori and colleagues performs SEM measurements *in situ* to investigate void anisotropy's relation to the anisotropic deformation response of polytetrafluoroethylene [27]. Another study from S. Fushimi and colleagues performs a similar analysis on polymeric foam materials [28]; however, these analyses on void evolution during deformation are only qualitative, and they do not quantify evolving void morphology throughout the bulk material.

While experimental studies have provided important insights about the mechanical performance of polymer materials, they are usually limited by the short length and timescales associated with structural changes in the material during deformation. Probing membrane mechanical properties on a molecular level, and frequently acquiring enough data on the changes of FVE during deformation, is challenging using experimental methods; Huang and colleges recently accomplished *in-situ* positron annihilation lifetime spectroscopy (PALS) experiments to observe free volume evolution in polyethylenes through active mechanical deformation on the minute scale [29], improving the throughput of this free volume analysis technique as well as circumventing chain relaxation effects over PALS's traditional, much longer length scales [30]–[32] . However, to gain a molecular-level insight about membrane mechanical response and the underlying physics on shorter timescales, molecular dynamics (MD) simulations provide a useful tool. MD simulations can elucidate valuable details about the mechanical deformation of polymer membranes, such as the changes in FVE distribution and polymer chains' rearrangement, on relevant length scales.

In this study, we implement molecular dynamics (MD) simulations to investigate *in silico* the mechanical response of three types of amorphous polymers: polymethylpentene (PMP), polystyrene (PS), and HAB-6FDA thermally rearranged polymer (TRP). We subject the three polymers to a controlled uniaxial deformation and relate the stress/strain response to the change in FVE distribution of the three structures, as changes in FVE distribution influence many important properties of the membrane module such as permeability, selectivity, and mechanical strength. Each polymer is deformed at different temperatures (100K, 300K, 500K) and strain rates. We examine the evolving morphological distribution of FVEs at defined intervals throughout deformation to show the impact of structure deformation on the pore microstructure.

**METHODS**

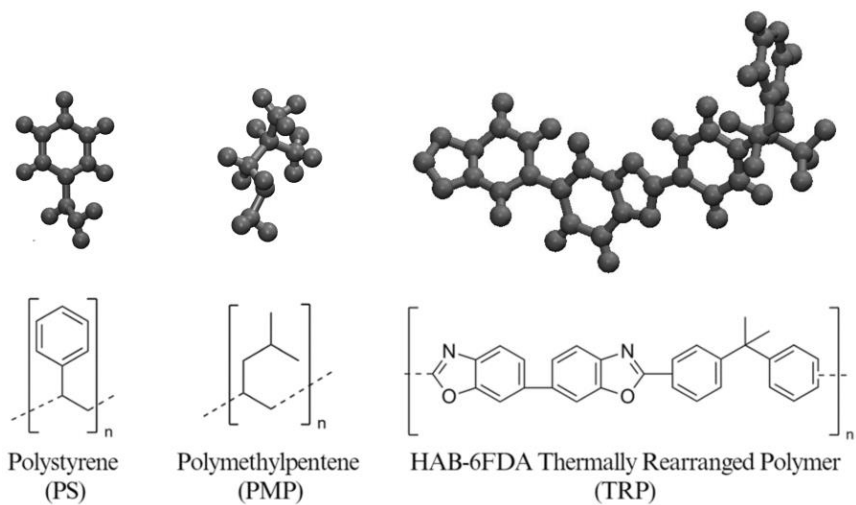

Figure 1 – (top) snapshots (bottom) chemical structures of the monomers used in this study.

**Equilibration**

In this study, we consider three polymers that represent different rigidities and microporosities: polymethylpentene (PMP), polystyrene (PS), and HAB-6FDA thermally rearranged polymer (TRP) (Figure 1). Details regarding the forcefield and system generation can be found in prior work. Briefly, polymers are modeled using the OPLS-AA forcefield parameters, obtained from the LibParGen web server [33], and 1.14 CM1A-LBCC charge model was used to calculate partial charges.

After parameterizing the monomer units, Polymatic [34], an open-source polymerization algorithm, was used to construct the polymer chains. Each polymer chain contains 200 monomers, and chains are terminated with hydrogen atoms. Energy minimization is performed after every polymerization, and a set of short simulations in the NVT and NPT ensembles are applied after every 3-to-5 polymerization steps allowing fragments to find each other. All the simulations were performed using the Large-scale Atomic/Molecular Massively Parallel Simulator (LAMMPS) [35] package with Verlet integrator and 1 fs timestep. Long-range interactions are calculated using a particle-particle particle mesh algorithm (PPPM) with a 1.5 nm cutoff. 20 of these chains are then packed to form the final system. After the chains are packed, we execute a 21-step equilibration process following the methods developed by Larson and Colina [36] to quench each of the three polymer systems at 1 atm to 100, 300, and 500 K (Table 1). Lastly, we run each system in the isothermal-isobaric NPT ensemble under these conditions for 5 ns and allow each system to fluctuate within the relaxed densities and temperatures.

Table 1 - Experimental and Modeled Densities of PMP, PS, and TRP

| Chemistry | Model $\rho$ at 100K (g/cm$^3$) | Model $\rho$ at 300K (g/cm$^3$) | Model $\rho$ at 500K (g/cm$^3$) | Experimental $\rho$ (g/cm$^3$) |
|---|---|---|---|---|
| PMP | 0.90 | 0.83 | 0.74 | 0.83 [37] |
| PS | 1.06 | 1.00 | 0.93 | 1.05 [38] |
| TRP | 1.34 | 1.30 | 1.22 | 1.36-1.45 [39] |

**Deformation**

The three polymer systems, PMP, PS, and TRP at 100 K, 300 K, and 500 K, are deformed at strain rates of 5 x $10^9$ (low), 25 x $10^9$ (medium), and 50 x $10^9$ (high) $s^{-1}$, informed by previous computational deformation studies of amorphous membranes [40]. To achieve this rate, uniaxial tensile deformation was applied along the x direction of the simulation box, with a constant engineering strain rate, by increasing the box dimension in finite steps. The extension ratio, or the ratio of current box length to the original length is varied at every MD time step. The barostat is applied in the y and z directions, to maintain an overall pressure of 1 atm. In LAMMPS, this translates to using a scale parameter of 2.0 (equivalent to 100% elongation) and simulation runtimes that resulted in the prescribed deformation rates—specifically, 400,000 femtoseconds (fs), 80,000 fs, and 40,000 fs for the low, medium and high strain rates, respectively. The axial component of the virial stress tensor ($\sigma_{xx}$) along the x direction is collected every 25 fs. Engineering strain is calculated by eq. 1.

$$\frac{l_o - l}{l_o} = \frac{\Delta l}{l_o} \qquad \text{(eq. 1)}$$

To create stress-strain curves, raw data generated from LAMMPS was averaged to reduce noise. This is achieved by block averaging $\sigma_{xx}$ over a width of 50 timesteps for 100K and 300K systems, and 100 timesteps for the 500K systems, except for PMP at 100 K and TRP at 300 K, for which 100-timestep window was applied. These windows were chosen so that the linear regime is separable across the 27 simulations, while also producing sufficiently smooth curves. The elastic modulus, also known as Young's modulus, is determined by calculating the slope of the linear regime of the stress-strain curve [41]. To remain consistent across the different systems, we chose to compute the modules using the lowest strain rate of 5 x $10^9$ $s^{-1}$, because of the smoothness of the linear regime at this strain rate. Three linear regressions were fit to data from the first smoothed data point (0.25% elongation for 100K & 300K simulations, and 0.50% elongation for 500K and PMP at 100K and TRP at 300K until the strain equaled 2%, 3%, and 4% elongation. The slopes of these regressions were averaged to optimize the linear fit and obtain a standard deviation from the average value.

**Free Volume Elements**

Quantitative studies of void evolution are conducted using the open-source package, Void Analysis Codes and Unix Utilities for Molecular Modeling and Simulation (VACUUMMS) [42]. This package employs the Cavity Energetic Sizing Algorithm (CESA), which is a Monte Carlo simulation-based technique for cavity sizing, to generate one million random cavity samples. CESA defines cavities as spherical volumes whose energy centers are the local minimum in a repulsive particle energy field. With CESA, we generate one million random cavity samples for each structure and then remove overlapping and identical cavity samples; this yields approximately 10,000 unique cavities in PMP and PS and approximately 25,000 unique cavities in TRP (reflective of TRP's larger chains and, subsequently, its larger box size). The distribution of spherical voids' diameters are plotted as histograms, and their visual representations are rendered from Scene Descriptive Language (SDL) by the open-source POV-ray tracing software. These FVE characterization techniques were executed on membranes at 100K, 300K, and 500K undergoing deformation at $25 \times 10^9$ s$^{-1}$ at three timesteps through the deformation period that correspond to 0%, 10%, and 20% elongation.

**RESULTS**

Uniaxial deformation is applied to PMP, PS and TRP at 100 K, 300 K, and 500 K at three different strain rates as described in the Methods section. The stress strain curves for PMP are shown in Figure 2. We see that for each temperature considered, the yield stress (maximum stress before plastic response sets in) increases with an increase in strain rate. This is common in polymers, where a high strain rate (high frequency) prevents the material from flowing, and low strain rates (low frequency) enables flow. As polymers are viscoelastic, time and temperature are equivalent, thus, effects of increasing temperature on the mechanical response are similar to the effects of decreasing the strain rate at lower temperatures, which we clearly observe. The elastic modulus (slope of the elastic regime) of PMP shows a similar trend to the yield stress, where it increases with strain rate and decreases with temperature, signifying that the material becomes stronger and stiffer under these conditions. At 100 K (Figure 2a), the response of PMP is that of a brittle plastic, however, as our model does not include breakable bonds, we will not see fracture. At 300 K (Figure 2b), PMP starts to display a response similar to a tough plastic, with a gradual decrease in stress post yielding. At 500 K (Figure 2c), it behaves as an elastomer, with yielding at high strains. Overall, reducing the temperature from 100 K to 500 K drastically changes the mechanical properties in PMP. Specifically, the yield stress at moderate strain rate increases by 200% from 500 K to 100K. This is indicative of the fact that the material changes state, from glassy to rubbery.

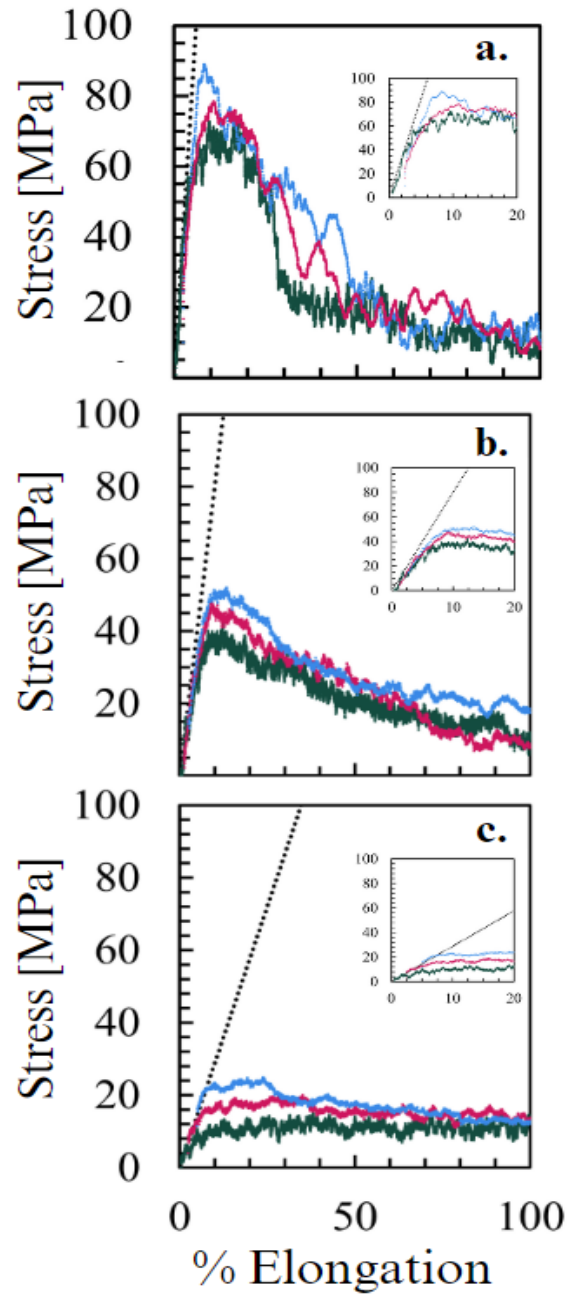

**Figure 2:** Engineering stress-strain curves for PMP at (a) 100K, (b) 300K, and (c) 500K. Each system is deformed at three strain rates: 5 x 10$^9$ s$^{-1}$ (green), 25 x 10$^9$ s$^{-1}$ (red), and 50 x 10$^9$ s$^{-1}$ (blue). Dashed lines are fit to the linear initial regime show the apparent differences in the elastic modulus. Inset shows the linear response regime.

The stress-strain curves for PS are shown in Figure 4. We observe the same temperature and strain rate effects for PS as PMP. However, at each temperature and strain rate, we see that PS has higher elastic modulus and yield stress than PMP. This is expected given that PS has phenyl sidechains, which impart higher strength to the polymer as it has the propensity to align through the formation of pi-pi bonds, unlike the methyl groups that belong to the PMP sidechain. We also observe that the stress-strain response of PS is similar at 100 K and 300 K (Figures 3a and 3b), as both these temperatures are below the $T_g$ of PS. At 500 K, beyond the $T_g$ of PS, we see a more elastomeric response. While PS also shows a decrease in mechanical properties from 100 K to 500 K, similar to PMP, it is not as steep as PMP's decrease, as evidenced by the changes in the slopes of the linear regimes (inset in Figure 3). The yield stress increases by 100% as the temperature is reduced from 500 K to 100 K.

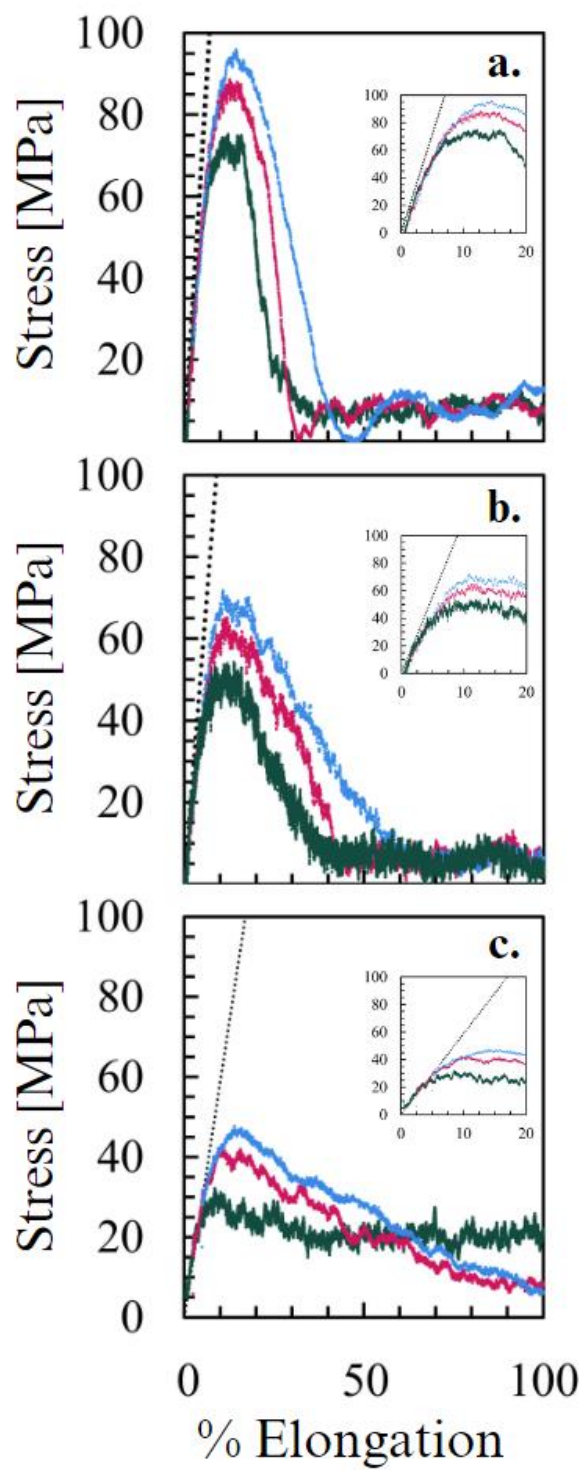

**Figure 3: Engineering stress-strain curves for PS at (a) 100K, (b) 300K, and (c) 500K. Each system is deformed at three strain rates: 5 x 10$^9$ s$^{-1}$ (green), 25 x 10$^9$ s$^{-1}$ (red), and 50 x 10$^9$ s$^{-1}$ (blue). Dashed lines are fit to the linear initial regime show the apparent differences in the elastic modulus. Inset shows the linear response regime**

As evidenced by the stress-strain curves (Figure 4), TRP follows the same dependence on temperature and strain rate as PMP and PS, as expected. However, the change in mechanical properties with temperature is the lowest for TRP compared to the other systems. This is evidenced by both its yield stress and elastic modulus being similar across the three temperatures. Particularly, the yield stress at moderate strain only increases by 50% as the temperature decreases from 500 K to 100 K, in stark contrast to PS and PMP. This is because TRP has the highest $T_g$ of the three polymers considered, and the three temperatures considered in this study are below the $T_g$ of TRP.

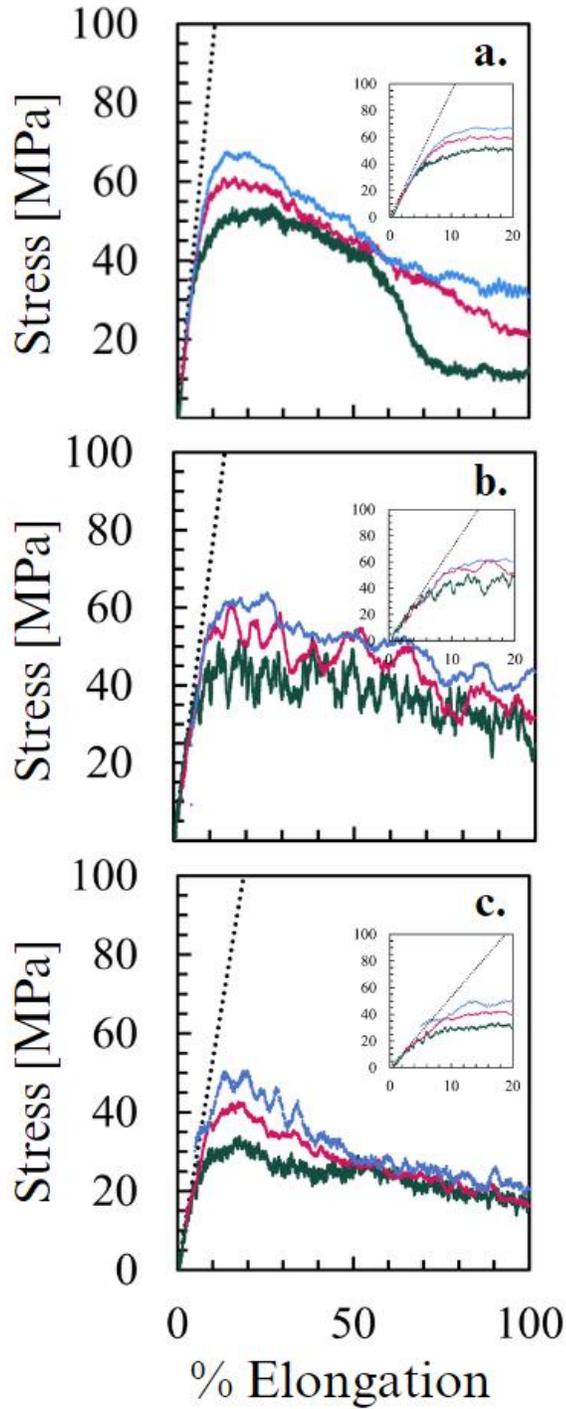

**Figure 4: Engineering stress-strain curves for TRP at (a) 100K, (b) 300K, and (c) 500K. Each system is deformed at three strain rates: 5 x 10$^9$ s$^{-1}$ (green), 25 x 10$^9$ s$^{-1}$ (red), and 50 x 10$^9$ s$^{-1}$ (blue). Dashed lines are fit to the linear initial regime show the apparent differences in the elastic modulus. Inset shows the linear response regime**

To better demonstrate change in mechanical properties, we compute the elastic modulus as described in the Methods section for the low strain rate of 5 x $10^9$ $s^{-1}$. As we do not include breakable bonds, the elastic linear response regime, which spans strains < 10%, can be used to compare to prior experiments at 300 K. We plot this for the three systems at different temperatures, as shown in Figure 5. In line with the yield stress, we observe that PMP has the most dramatic reduction in the value of its elastic modulus as temperature increases, followed by PS, and finally TRP. A higher elastic modulus implies that the material is stiff, and resists stretching. Specifically, the modulus increases by 466%, 133% and 64% with a decrease in temperature from 500 K to 100K for PMP, PS, and TRP, respectively. This stems from the fact that 500 K is above the glass transition for both PMP and PS, but below the $T_g$ for TRP. Interestingly, the trends across the three systems are different at different temperatures. At 100 K, PMP has the highest modulus, followed by PS, and finally TRP. In prior work, we observe that at 100 K, PMP had several small voids, with a much narrow distribution compared to PS and TRP. Overall, as porosity increases, modulus decreases, as seen in prior experimental work as well. On the other end of the temperature spectrum, at 500 K, PS and TRP have similar moduli of approximately 6 MPa, and the modulus of PMP is approximately 3 MPa, which we attribute to the elastomeric nature of PMP. At 300 K, we find that PS has the highest modulus (11.5 MPa), with PMP and TRP having comparable moduli (8.2 and 7.9 MPa, respectively). At this temperature, TRP has a broader distribution of voids than the other systems, however, the mean void size is comparable across the three systems (~ 2 angstroms). Another contributor to the increased modulus of PS at 300 K could be that it is closer to its equilibrium density than TRP. Owing to its high free volume at temperatures below $T_g$, the density of TRP is vastly different from its equilibrium density. This contributes to TRP's high rate of aging (densification). In terms of their thermodynamic free energy landscapes, PS is closer to its global energy minimum compared to TRP. Therefore, higher values of stress are required to bring PS out of this energy minimum, which translates to a higher elastic moduli and yield stress at a given temperature. Additionally, at 300 K, the chains have sufficient mobility that they are able to orient along the direction of pull, which contributes to the strength of the material.

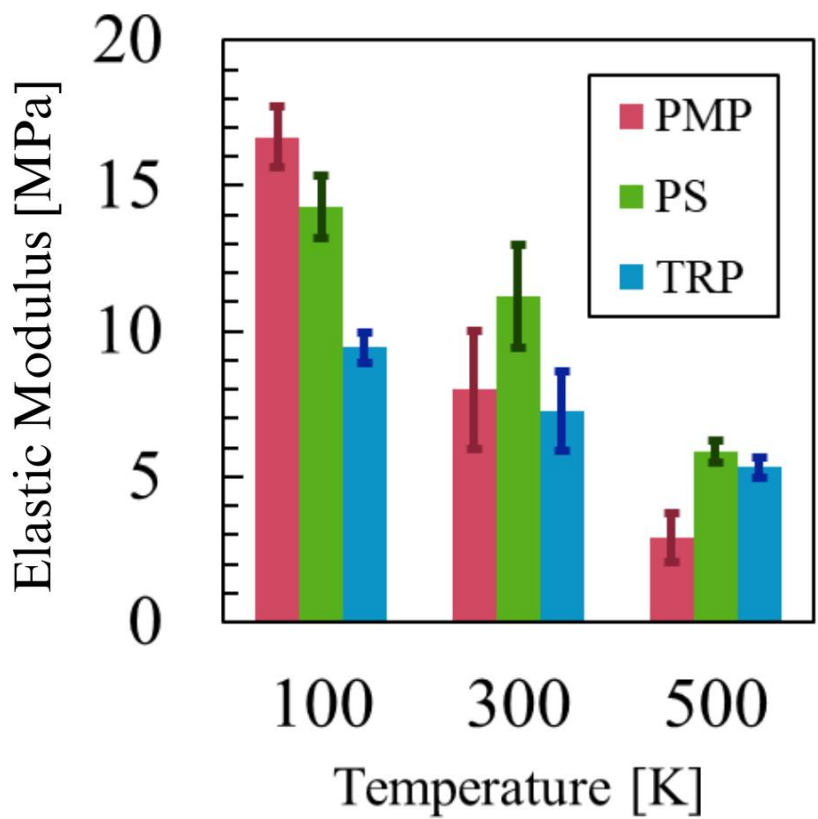

**Figure 5: Elastic moduli for PMP (red), PS (green), and TRP (blue) at each deformation temperature. Data sourced from deformations at strain rate of 5 x $10^9$ s$^{-1}$. Error bars are representative of a standard deviation from the average moduli values, calculated from three samples of each linear response regime.**

To investigate the phenomenon of polymer orientation further, we capture snapshots of the three systems at 300 K at different strains (Figure 6). We see that at moderate strains (20%), both TRP and PMP do not show the onset of voids or cavitation (Figure 6a and 6c). However, at the same strain, we see that PS starts to cavitate. We reason that the high elastic modulus of PS at 300 K and the formation of cavities in the PS matrix well before TRP and PMP are indicative of the fact that the chains of PS have a greater propensity to be oriented upon pull, due to the phenyl sidechain groups' ability to maintain long ranged order. In contrast, TRP has several cyclic groups along the backbone, which do not orient as well as PS due to the rigidity of the chain. PMP chains, with its methyl side groups, are not likely to orient upon strain. Additionally, from the stress-strain curves of the materials at 300 K, there is a much-steeper decline following the yield stress near 20% elongation for PS (Figure 3b), compared to PMP and TRP (Figures 2b and 4b), which may be attributed to the void formation seen in Figure 6b. While we do not incorporate breakable bonds in our model, we reason that, in the absence of fracture, high strain behavior can provide insight into chain orientation. At 100% elongation (Figure 6) we see distinct linearization of chains in PS, with the propagation of voids and formation of fibrils. In comparison, TRP and PMP do not show the formation of a few fibrils holding the system together, and both these systems display moderate number of voids, similar to that in the PS system at 20% elongation. Interestingly, we see the phenyl groups in the PS clearly oriented perpendicular to the direction of pull, with the backbone aligned parallel. While we see the orientation of the backbone in TRP and PMP as well, it is not as prominent as the chain alignment in PS. The trends in the elastic moduli for PS and TRP agree with experimental results, where Freeman et al. report the modulus of TRP to 2.51 GPa [43], and Harper et al. report the modulus of PS to be 3.1 GPa. We note that these are different studies, and the conditions under which these measurements were performed might not be consistent, confounding direct comparisons. Nevertheless, we report these values, as there are not many experiments characterizing TRP.

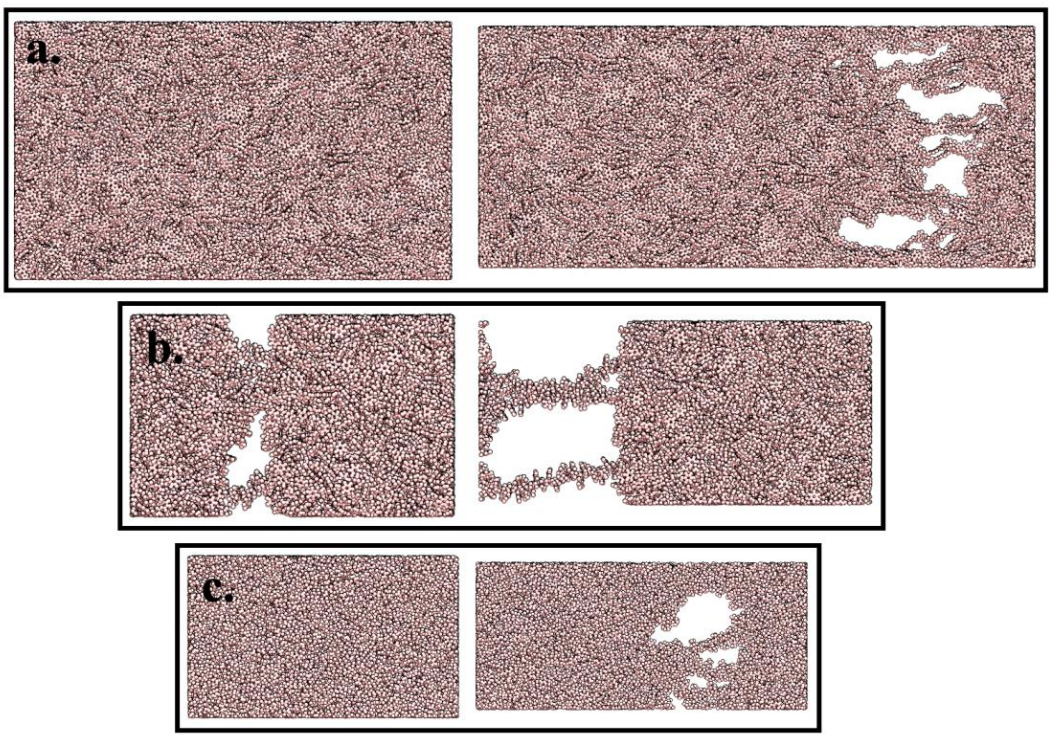

**Figure 6: Chain visualization of TRP (a.), PS (b.), and PMP (c.) modules, captured with the Visual Molecular Dynamics (VMD) package, at 20% (left) and 100% elongation (right). Data sourced from deformations at a rate of v and 300K.**

Lastly, we characterize the change in free volume distribution with deformation. FVEs are microstructural features of amorphous polymers, and quantifying their evolution can provide insight into the mechanical behavior of each material. For the systems considered, we show the changes in FVEs at the extreme temperatures (100 K and 500 K) at three distinct strain points – 0%, 10% and 20% elongation for a strain rate of $25 \times 10^9$ s$^{-1}$. We chose these strain rates as they represent three distinct states – the initial state, linear response regime, and near the yield point. We also show quantitative snapshots using POV-ray and SDL renderings of CESA-generated cavities, at points of 0% and 20% elongation. Overall, we see that the free volume shifts to broader distributions with an increase in strain, with a decrease in the smaller pores. Figure 7 shows the free volume evolution for PMP. The change in free volume is greater at 100 K compared to 500 K. This is also seen visually in the snapshots – the size of the voids at 20% is significantly greater than that at 0% for 100 K, compared to 500 K, where we do not see a prominent change in voids. The inset of Figure 7 shows a clear shift of the distribution to the right, signifying that, during deformation, small voids come together to become larger voids, as it is accompanied by a decrease in total number of smaller cavities. At 500 K, there is only a small difference in the distribution between 10% and 20% elongation. This might be because the material shows an elastomeric tensile response, where microstructure might not change very much after a certain strain.

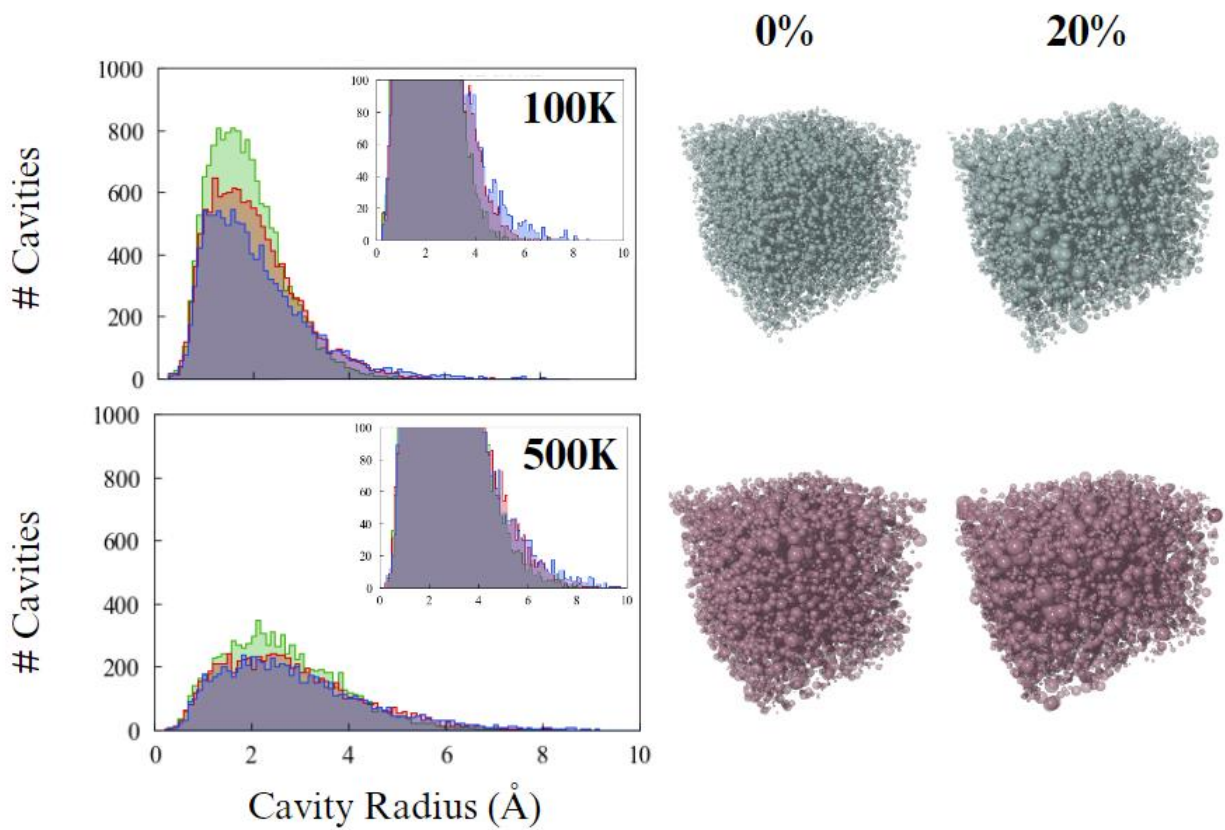

**Figure 7: PMP's FVE radius distribution evolving through deformation from 0% (green) to 20% (blue) elongation. Red represents 10% elongation. Data is collected with VACUUMMS from membranes deforming at strain rate of 25 x 10$^9$ s$^{-1}$ at 100K and 500K.**

We see a similar trend in PS, where the void size increases with increasing strain, accompanied by an overall decrease in smaller voids. At 100 K and 20% elongation, we see a single large void at 9 Ang, which indicates the formation of cavities. This is in line with the PS snapshots in

Figure 5. At 500 K there is a considerable difference between the distributions at 10% and 20% elongation, unlike PMP, signifying that the PS's microstructure is considerably different at low and moderate strains, even at high temperatures.

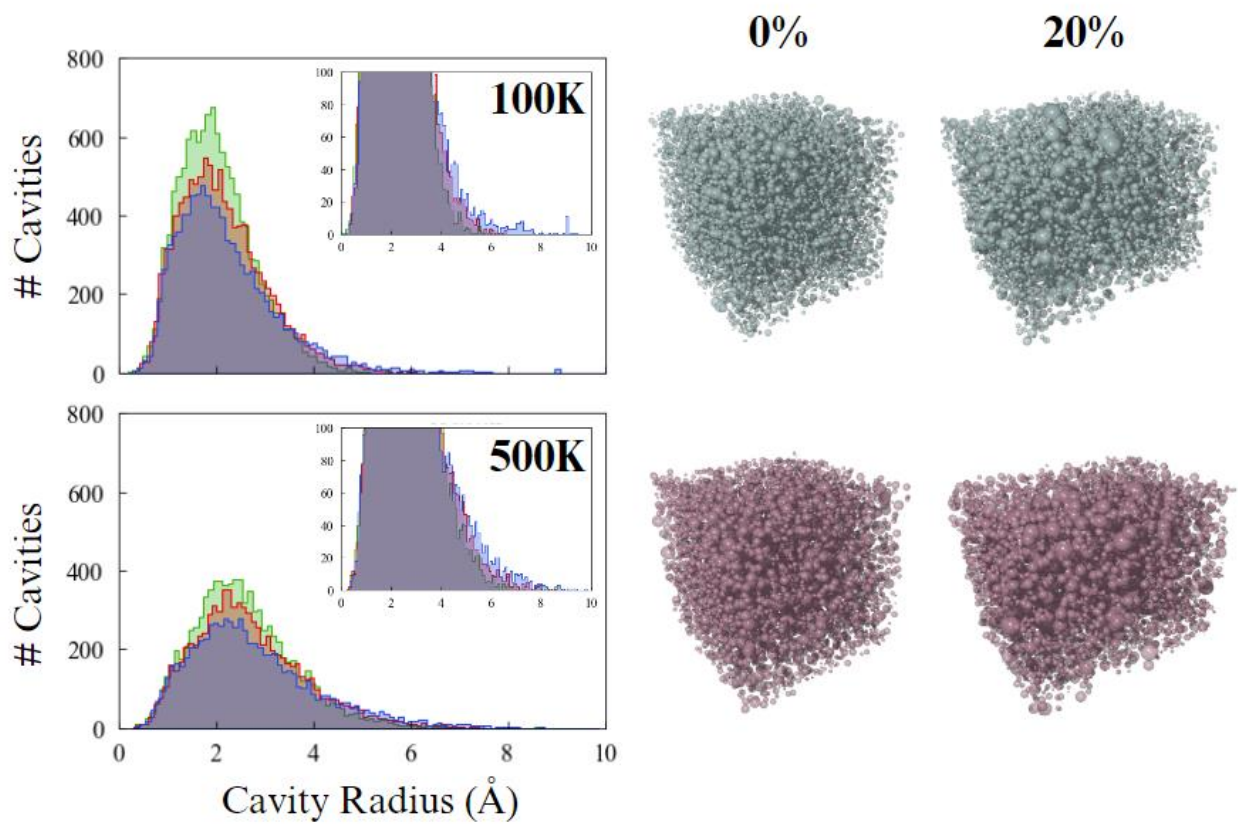

**Figure 8:** PS's FVE radius distribution evolving through deformation from 0% (green) to 20% (blue) elongation. Red represents 10% elongation. Data is collected with VACUUMMS from membranes deforming at strain rate = 25 x $10^9$ $s^{-1}$ at 100K and 500K.

At 100 K, TRP shows an increase in large voids at the expense of smaller ones, as we observe with the other systems, however, this change is not as prominent as it is in either PS or PMP. This might indicate why TRP also has a lower modulus at this temperature (Figure 5); the microstructure does not change as dramatically to accommodate large stresses brought about by deformation, as compared to the other systems. At 500 K, in line with PS, there is a substantial change in the voids at 10% compared to 20%. We note that TRP's microstructure features voids beyond 10 Å, which we do not observe in either PS or PMP, as seen in Supporting Information.

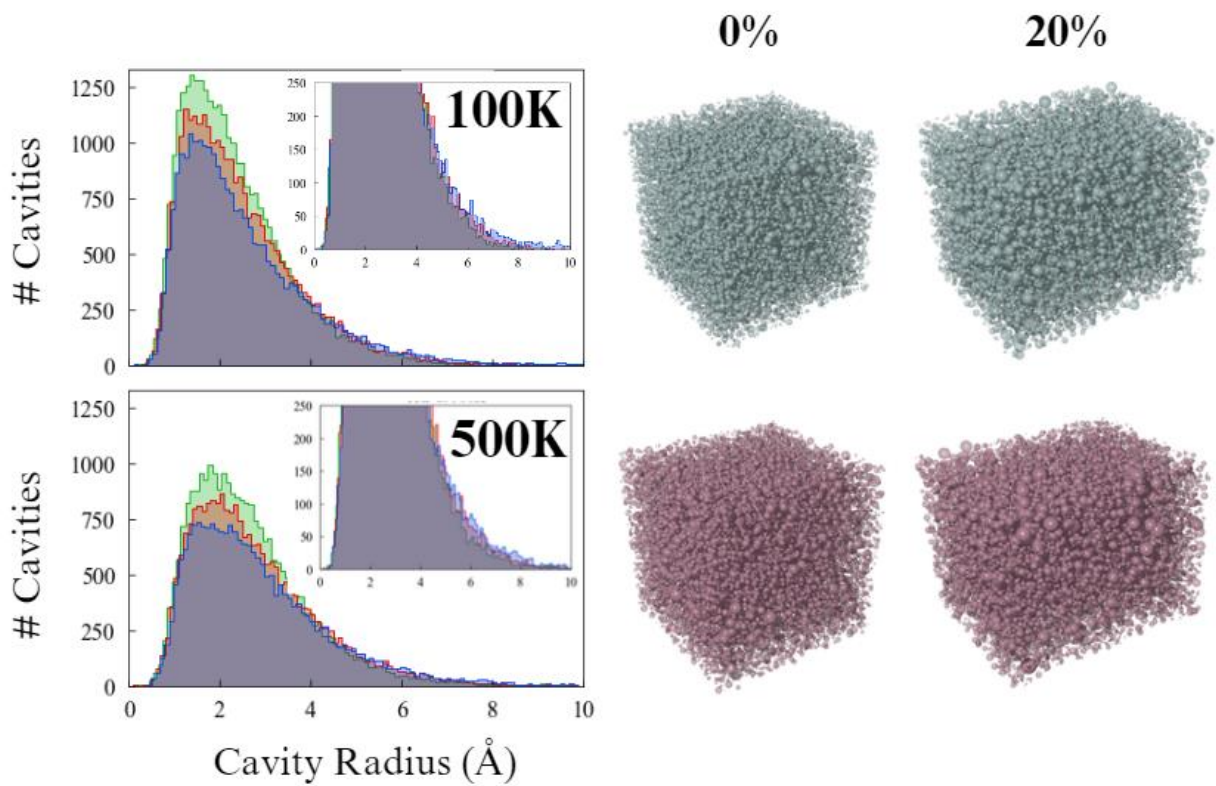

Figure 9: TRP's FVE radius distribution evolving through deformation from 0% (green) to 20% (blue) elongation. Red represents 10% elongation. Data is collected with VACUUMMS from membranes deforming at strain rate = 25 x 10$^9$ s$^{-1}$ at 100K and 500K.

Overall, the FVE data reveal several relationships between polymer rigidity and FVE robustness. In prior work, we see that increasing temperature widens the distribution of FVE diameters and correlates with the glass transition temperature of these materials. PMP has the least stable FVE distribution, followed by PS, finally TRP exhibits the least change. Second, in PMP and PS, large voids form later during the deformation period. This is attributed to the small pores coming together as the polymer deforms, with a decrease in the frequencies of voids with small-to-medium diameters and the appearance of new voids at higher diameter. Lastly, we find that TRP exhibits the most rigid FVE distribution, shown by its resistance to change not only through deformation but also across temperatures.

**Conclusion**

To elucidate the relationship between membrane chemistry, membrane flexibility, and the evolution of void morphology throughout membrane deformation, we utilize all-atom MD simulations to model uniaxial deformation of membrane materials made of PMP, PS, and TRP at 100K, 300K, and 500K at 1 atm. These structures are constructed using the simulated polymerization algorithm Polymatic and the interatomic interactions are modeled using OPLS-AA force field. The three polymers are uniaxially deformed at three strain rates, and the evolving FVE distributions in each deforming membrane are studied using both qualitative and quantitative measures on membranes deforming at moderate strain.

Our results show that higher strain rates and lower temperatures promote higher values of a membrane's yield stress across all simulations. When comparing the yield stress values of each membrane at a specific temperature, PS has the highest value of the membranes across all three temperatures, which we attribute to both the chain alignment and void evolution.

TRP's FVE distribution displays significant robustness throughout the deformation period across different temperatures. This data promotes the expectation that TRP will maintain its designed sieving performance throughout deformation, irrespective of temperature, so long as the membrane is kept below its $T_g$. Our findings underscore the importance of balancing membrane mechanical properties with the evolution FVE distribution for membrane separation applications.

In future work, we aim to investigate the linearization and dynamics of the polymer chains throughout the deformation period to determine whether the improved modulus of PS can be attributed to the presence of cyclic pendant groups along its backbone.

**Acknowledgments**

This work was supported by Dr. Janani Sampath's startup funds, provided by the University of Florida Department of Chemical Engineering. Financial support was further provided by the University of Florida Center for Undergraduate Research's University Scholars Program. Computational techniques were executed using HiPerGator 3.0, hosted by the University of Florida Research Computing. The authors would like to acknowledge the contributions of Dr. Frank Willmore for his development of the open-source VACUUMMS package, mentorship, and troubleshooting advice.